\documentclass{article}
\usepackage{epsfig}
\usepackage{amsmath}
\usepackage{amsfonts}
\usepackage{amssymb}
\usepackage{euscript}
\usepackage{graphicx}
\usepackage{multicol}
\usepackage{lscape}
\usepackage{pstricks}
\usepackage{multirow}

\usepackage[cp1251]{inputenc}
\usepackage[russian]{babel}

\begin{document}
\sloppy

\begin{center}
{{\Large\bfseries Should the electric polarization of intrastellar plasma be taken into consideration at elaboration a star theory?
\\}}
\vspace{.3cm}
B.V.Vasiliev
\end{center}

The gravity-induced electric polarization of the "ordinary" atomic substances such as  gases, liquids, solids is negligible and plays no role in the forming of their equilibrium state. In metals, it is very small too {\cite{Leung}},{\cite{BV-Vir}}. By inertia, the value of gravity-induced electric polarization in the hot dense plasma of the star interior is postulated as small and it has never been taken into consideration.

However, the methodology of determining of correctness of the postulates is known since Galileo. According to this methodology, a theoretical consideration of the phenomenon must be made on a base of a postulate, following by comparison of the correlation of the theory to experimental data.\\

In the particular case of the consideration of the electric polarization of intrastellar plasma, two contradictory postulates could be made:\\

 \hspace{1cm}1) The electric polarization of intrastellar plasma should not be taken into consideration for modeling a star theory.\\

 \hspace{1cm}2) It is necessary to take it into consideration.\\

 The theory created on the basis of the first postulate is the physics of stars in its present form. However, modern astrophysics does not make a comparison of the theory with measurement data. For the main reason, these measurement have not been fulfilled until recently. 
 As a result, the modern theory of stars is in a state of  a pre-Galileo science built on the basis of dogmata. The founders of
astrophysics have created a model of stars, while other considerations assumed to be correct if their results coincide with this model, and wrong if they contradict it.

And what about the measurements?

To the present, astronomers have measured {\cite{BV-arXiv}}:\\

 \hspace{1cm}       the  distribution of the stars' mass,\\

\hspace{1cm}        the correlation of stellar temperature to its mass, \\

\hspace{1cm}        the  correlation  of stellar radii and its mass,\\

\hspace{1cm}        the spectrum of seismic oscillation of the Sun,\\

 \hspace{1cm}       the correlation of stellar magnetic field to stellar rotation velocity,\\

\hspace{1cm}    the velocity of periastron rotation of closed binary stars.\\

There are some other correlations related to the listed above. For example, the correlations of the stellar luminosity to its mass, discovered about a hundred years ago.
By the modern theory of stars, these correlations can not be quantitatively explained, and are considered to be empirical(?). Majority of astrophysicists simply do not take them into attention, being satisfied by qualitative explanations.

Therefore, the situation is quite dramatic. The absence of quantitative agreement between conclusions of the theory of stars and the observations may be explained by one reason only: the existing theory based on the postulate of no importance of the electric polarization of the intrastellar plasma can not correctly describe the internal structure of stars.\\

In contrast, the theory of stellar structure {\cite{BV-arXiv}} designed through taking the gravity-induced electric polarization of intrastellar plasma into account, describes their characteristic features. According to this theory, the traits of the variety of stars can be characterized by their chemical compositions, as they depend on two parameters only: A and Z, the mass number and the charge number of atomic nuclei composing intrastellar plasma. It was observed that the existence of stars with an integer ratio $A/Z=3;4;5…$ is the most probable. The heaviest stars must have $A/Z=1$, and the mass about 25 times the mass of the Sun, which is in very precise agreement with the observations.
 Another characteristic  feature of the stars determined via of the electric polarization, is that they have cores with a high fixed density and temperature.
 The existence of core excludes the possibility of the collapse of stars.
This theory can explain {\it{all}} known now correlations of  stellar parameters quantitatively  and often with significantly high accuracy. On the whole, comparing its results with the measurement data fully confirms entirely the importance of the electric polarization of intrastellar plasma in achieving its equilibrium state.\\

However, this theory also has its difficulties. For example, it is unclear how the existence of the solar neutrino flux can be explained through this theory.\\

In other words, the answer to the title question is: of course, the gravity-induced electric polarization of intrastellar plasma must be taken into consideration at formulation of a star theory, as plasma in the stars is electrically polarized.

\end{document}